\documentclass[submitting]{nst}

\usepackage{subfigure,dcolumn}
\usepackage[T2A,T1]{fontenc}
\usepackage[russian,english]{babel}
\usepackage{multirow}
\fancyhfoffset{0pt}

\usepackage{listings}
\begin{document}

\title{Inferring Nuclear Radial Parameters from Initial-State Information in Heavy-Ion Collisions Using Deep Neural Networks}

\thanks{This work is supported by the National Natural Science Foundation of China (NSFC) under Grant Nos.~12535010 and 92570117, the Ministry of Science and Technology of China under Grant No.~2024YFA1611004, the Shenzhen Peacock Fund under Grant No.~2023TC0179, the CUHK-Shenzhen University Development Fund under Grant Nos.~UDF01003041 and UDF03003041, the China Scholarship Council (CSC) under Grant Nos.~202406770003 and 202406770090, and the China Postdoctoral Science Foundation under Grant No.~2025M781519.}

\author{Yang Liu}
\affiliation{Key Laboratory of Quark \& Lepton Physics (MOE) and Institute of Particle Physics, Central China Normal University, Wuhan 430079, China}

\author{Yu Sha}
\affiliation{School of Science and Engineering, The Chinese University of Hong Kong, Shenzhen (CUHK-Shenzhen), Guangdong, 518172, China}

\author{Xiang Fan}
\affiliation{Key Laboratory of Quark \& Lepton Physics (MOE) and Institute of Particle Physics, Central China Normal University, Wuhan 430079, China}

\author{Yan-Sheng Tu}
\affiliation{School of Science and Engineering, The Chinese University of Hong Kong, Shenzhen (CUHK-Shenzhen), Guangdong, 518172, China}

\author{Shi-Lei Tao}
\affiliation{Key Laboratory of Quark \& Lepton Physics (MOE) and Institute of Particle Physics, Central China Normal University, Wuhan 430079, China}
\affiliation{Helmholtz Research Academy Hesse for FAIR (HFHF), GSI Helmholtz Center for Heavy Ion Research, Frankfurt am Main 60438, Germany}
\affiliation{Institute for Theoretical Physics, Johann Wolfgang Goethe University, Frankfurt am Main 60438, Germany}

\author{Qing Zhang}
\affiliation{China Nuclear Power Operation Technology Corporation, China National Nuclear Corporation, Wuhan 430223, China}

\author{Jun-Qi Tao}
\email[]{taojunqi@mails.ccnu.edu.cn}
\affiliation{Key Laboratory of Quark \& Lepton Physics (MOE) and Institute of Particle Physics, Central China Normal University, Wuhan 430079, China}
\affiliation{School of Science and Engineering, The Chinese University of Hong Kong, Shenzhen (CUHK-Shenzhen), Guangdong, 518172, China}
\affiliation{Helmholtz Research Academy Hesse for FAIR (HFHF), GSI Helmholtz Center for Heavy Ion Research, Frankfurt am Main 60438, Germany}
\affiliation{Institute for Theoretical Physics, Johann Wolfgang Goethe University, Frankfurt am Main 60438, Germany}

\author{Kai Zhou}
\email[]{zhoukai@cuhk.edu.cn}
\affiliation{School of Science and Engineering, The Chinese University of Hong Kong, Shenzhen (CUHK-Shenzhen), Guangdong, 518172, China}
\affiliation{School of Artificial Intelligence, The Chinese University of Hong Kong, Shenzhen (CUHK-Shenzhen), Guangdong, 518172, China}

\author{Hua Zheng}
\email[]{zhengh@snnu.edu.cn}
\affiliation{School of Physics and Information Technology, Shaanxi Normal University, Xi'an 710119, China}

\author{Ben-Wei Zhang}
\email[]{bwzhang@mail.ccnu.edu.cn}
\affiliation{Key Laboratory of Quark \& Lepton Physics (MOE) and Institute of Particle Physics, Central China Normal University, Wuhan 430079, China}

\begin{abstract}
The geometric structure of atomic nuclei plays an important role in shaping the initial conditions of high-energy nuclear collisions and the subsequent evolution of the quark--gluon plasma. In this work, we develop a multimodal deep-learning framework to infer the Woods--Saxon radius ($R_0$) and diffuseness parameter ($a$) from initial-state information. Using the TRENTo model for relativistic $^{238}$U+$^{238}$U collisions, we construct a dataset covering multiple combinations of $R_0$ and $a$, and perform parameter regression by combining two-dimensional entropy-density distributions with event-level scalar observables, including multiplicity and eccentricities $\varepsilon_2$--$\varepsilon_5$. The results show that the proposed framework can effectively recover the Woods--Saxon parameters from initial-state information. Joint multi-event inputs suppress statistical fluctuations and improve regression performance by enhancing the extraction of collective information. The inversion accuracy exhibits a pronounced centrality dependence: $R_0$ is best constrained in central collisions, whereas $a$ shows the strongest sensitivity in the mid-peripheral region and becomes less well constrained in ultra-central and peripheral collisions. These findings indicate that the initial-state geometry retains useful information about nuclear radial structure and provide a baseline for future studies that incorporate transport evolution and experimental final-state observables.
\end{abstract}


\maketitle
\nolinenumbers

\section{Introduction}
 
Relativistic heavy-ion collisions offer a unique opportunity to study strongly interacting QCD matter under extreme conditions~\cite{BraunMunzingerStachel2007,Shuryak2017SCQGP,GyulassyMcLerran2005NewFormsQCD}, creating the quark-gluon plasma (QGP) at high energies. The properties of the QGP are inferred from final-state observables~\cite{Shuryak2017SCQGP, GaleJeonSchenke2013}. It is well established that these observables are shaped not only by the subsequent medium evolution, but also by the geometry and event-by-event fluctuations of the initial state~\cite{LuzumPetersen2014,AlverRoland2010TriangularFlow,Petersen2013InitialFluctuations,ZhuWuQin2025FlowDecorrelation}. Understanding the structure of the colliding nuclei is therefore an important issue nowadays in heavy-ion physics~\cite{Jia2024ImagingInitialCondition,ALICE2025NuclearStructure}.

Traditional probes such as elastic electron scattering and Coulomb excitation provide important information on nuclear structure~\cite{deVriesdeJagerdeVries1987ChargeDensity,AngeliMarinova2013ChargeRadii,Cline1986CoulombExcitation,Raghavan1989NuclearMoments}. In addition, heavy-ion collisions may provide complementary access to nuclear-structure effects through final-state observables that are sensitive to the initial geometry and its fluctuations~\cite{LuzumPetersen2014,Petersen2013InitialFluctuations,Jia2024ImagingInitialCondition,GiacaloneJiaZhang2021Deformation,JiaZhang2023ScalingStructure,Giacalone2024U238Shape,Cheng:2023ucp}. For example, anisotropic flow observables provide a useful probe of nuclear deformation through their sensitivity to the initial eccentricity of the colliding system~\cite{GiacaloneJiaZhang2021Deformation,LuZhaoLi2023AnisotropicFlow,JiaHuangZhang2022Quadrupole,ATLAS2023FlowPtXePb}. Neutron skin can be explored through observables associated with the participant--spectator geometry and the neutron/proton distribution near the nuclear surface, such as free spectator nucleon yields and the free spectator neutron-to-proton ratio $N_n/N_p$~\cite{DingFangMa2024NeutronSkin,JiaXuZhang2020NeutronSkin,LiuZhangXu2021IsobarSkinShape,JiaGiacaloneZhang2023SkinDeformation}. Cluster structure may be accessed through observables sensitive to initial spatial correlations and event-by-event fluctuations, including flow coefficients, multiparticle correlations, and certain spectator-related observables~\cite{BroniowskiRuizArriola2014AlphaClusterSignatures,BozekBroniowskiRuizArriolaRybczynski2014AlphaClustersFlow,MaMaZhang2020AnisotropyFluctuation}. However, as important baseline, comparatively little attention has been devoted to the identification of Woods--Saxon parameters $R_0$ and $a$, which characterize the radial distribution of the nucleus. These parameters can affect the overlap geometry, entropy deposition, and final-state observables~\cite{Loizides2015PHOBOSGlauber,Jia2022ShapeAtomicNuclei,Giacalone2023NeutronSkinPb208}. In particular, AMPT model studies of Xe+Xe collisions have shown that the diffuseness parameter $a$ can have a significant impact on $v_2$ in mid-central collisions~\cite{LuZhaoLi2023AnisotropicFlow,PandeySingh2024XeDeformation}. This makes it important to determine whether, and how, information on the Woods--Saxon radial parameters is encoded in the collision evolution.

Recent advances in neural-network methods have greatly improved our ability to extract useful information from complex and high-dimensional data, especially for nonlinear problems in which the relevant information is weak, distributed across correlated inputs, and difficult to identify through simple analyses~\cite{Schmidhuber2015DeepLearningOverview,BarkerBuhmannCarleoEtAl2022MLNP,BrehmerLouppePavezCranmer2020LearningHEP}. More broadly, machine-learning methods have been widely used in nuclear and high-energy physics~\cite{Zhou:2023pti, HeMaZhou2023MLReviewNST,PangZhouWang2021MLHeavyIon}, including applications to the equation of state~\cite{PangZhouSu2018EOSMeter,PangZhouWang2021MLHeavyIon,OmanaKuttan:2020btb}, event classification~\cite{YangFanLi2024OrientingEvents,HeHeMa2021InitialClusteringML}, jet tomography~\cite{DuPablosTywoniuk2021JetModificationDL,DuPablosTywoniuk2022JetTomographyDL}, flow analysis~\cite{HirvonenEskolaNiemi2023FlowDL}, data-driven prediction and interpolation of bulk observables across collision energies and centralities~\cite{TaoFanLiuEtAl2026MultiObservable}, chiral-magnetic effects~\cite{ZhaoWangZhou2022CMEDL,GuoWangZhou2025CMEUnfolding}, and structure-related information extraction in heavy-ion collisions~\cite{SahaMehrabpourJia2025NuclearStructureNN}. 

In this work, we use the TRENTo model~\cite{MorelandBernhardBass2015TRENTo} to generate initial-state entropy-density profiles for relativistic $^{238}\mathrm{U}+^{238}\mathrm{U}$ collisions over multiple combinations of the Woods--Saxon parameters $R_0$ and $a$, and develop a multimodal deep-learning framework to infer these parameters from the generated initial-state information. The network combines two-dimensional entropy-density distributions with event-level scalar observables, including the multiplicity and eccentricities $\varepsilon_2$--$\varepsilon_5$, thereby incorporating both spatial and fluctuation-sensitive information into the regression. We examine how the regression performance depends on the number of events analyzed jointly and on collision centrality. We find that multi-event inputs improve the recovery of both parameters, while their centrality dependences differ: $R_0$ is best constrained in central collisions, whereas the regression performance for $a$ is higher across a broad mid-peripheral range than toward the central and peripheral ends. By restricting the present analysis to the TRENTo initial state, this study provides a baseline for future parameter inference using dynamically evolved final-state observables.

The remainder of this paper is organized as follows. In Section~II, we describe the Woods--Saxon parameterization, dataset construction, and network architecture. Section~III presents the regression results for $R_0$ and $a$ and discusses their dependence on event-group size and centrality. Finally, Section~IV summarizes the main conclusions and outlines possible future directions.

\section{METHOD}

In this study, the $^{238}\mathrm{U}$ nucleus is modeled using a deformed Woods--Saxon density distribution, parameterized as:
\begin{equation}
\label{eq:ws_density}
\rho(r,\theta) = \frac{\rho_0}{1 + \exp\left[\frac{r - R(\theta)}{a}\right]},
\end{equation}
where $R(\theta) = R_0\left(1 + \beta_2 Y_{20}(\theta) + \beta_4 Y_{40}(\theta)\right)$. Here, $\rho_0$ is a normalization constant, $R_0$ is the nuclear radius parameter, and $a$ is the diffuseness parameter. The deformation parameters $\beta_2$ and $\beta_4$ describe the quadrupole and hexadecapole deformations, respectively, and $Y_{20}(\theta)$ and $Y_{40}(\theta)$ are the corresponding spherical harmonics. In this work, $\beta_2 = 0.28$ and $\beta_4 = 0.093$ are fixed~\cite{FilipLednickyMasuiXu2009,SchenkeShenTribedy2020}, while the focus is on the impact of the radial structure parameters $R_0$ and $a$ on the initial state of heavy-ion collisions.

In the TRENTo event generation, the two uranium nuclei are randomly oriented. The tip--tip, tip--body, and body--body configurations are used only as representative fixed-orientation cases for the geometric analysis and are not treated as separate regression classes.

The training data for this model are generated through a two-dimensional parameter scan, where $R_0$ is sampled at 21 points from 6.0 to 7.0 fm, and $a$ is sampled at 21 points from 0.4 to 0.6 fm. This results in 441 parameter combinations and a total of $8.82 \times 10^6$ generated collision events. The data are randomly divided into training and validation sets with an 8:2 ratio. For model evaluation, an independent test set is generated from a denser parameter grid, with both $R_0$ and $a$ sampled at 41 points. For each centrality selection, this yields a total of 336,200 events. In all datasets, centrality is defined by the multiplicity obtained from the TRENTo-generated events, with larger multiplicity corresponding to more central collisions.

\begin{figure*}[!htb]
  \includegraphics[width=0.9\hsize]{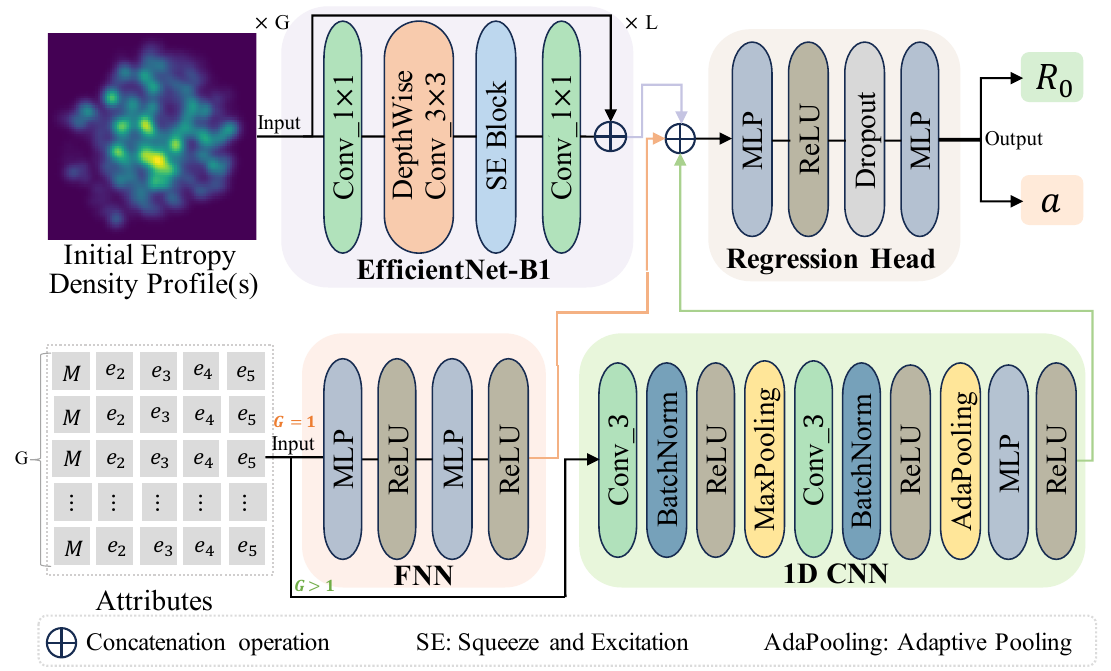}
  \caption{Multimodal neural-network architecture for the inversion of the Woods--Saxon radius $R_0$ and diffuseness parameter $a$ from heavy-ion initial-state information.}
  \label{fig:network}
\end{figure*}

A multimodal fusion regression network is utilized to predict the Woods--Saxon parameters $R_0$ and $a$ from the generated data, as shown in Fig.~\ref{fig:network}. The network consists of four main components. The image branch takes the two-dimensional initial entropy density distribution as input, represented by a tensor of shape $(B,G,H,W)$, where $B=64$ is the batch size, $G$ is the number of events in each input group, and $H=100$ and $W=100$ denote the image height and width, respectively. A pretrained EfficientNet-B1 is adopted as the backbone to extract spatial features from the initial-state images. Through stacked MBConv blocks and the squeeze-and-excitation mechanism~\cite{TanLe2019EfficientNet}, the network captures high-level spatial representations. After global average pooling, a 1280-dimensional image feature vector is obtained.
The attribute branch processes the event-level scalar inputs, organized as a tensor of shape $(B,G,L)$, where $L=5$ corresponds to the five scalar observables of each event, namely the multiplicity and eccentricities $\varepsilon_2$--$\varepsilon_5$. For the multi-event case ($G>1$), a one-dimensional convolutional neural network is used to encode inter-event correlations. The input tensor is first reshaped from $(B,G,L)$ to $(B,L,G)$ and then passed through two consecutive one-dimensional convolutional blocks. The first block maps the channel dimension from $L$ to 64, followed by batch normalization, ReLU activation, and max-pooling. The second block further maps the channel dimension from 64 to 128, followed by batch normalization and ReLU activation. The event-group dimension is then aggregated by adaptive average pooling, and a 64-dimensional attribute feature is obtained through a linear projection. For the single-event case ($G=1$), where inter-event correlations are absent, the attribute branch is simplified to a feedforward neural network. The input features are passed through two fully connected layers of dimensions $L \rightarrow 128 \rightarrow 64$, with a ReLU activation after each layer, yielding a 64-dimensional attribute feature.
In the fusion stage, the 1280-dimensional image feature and the 64-dimensional attribute feature are concatenated to form a 1344-dimensional joint representation, which is then fed into the regression head. The regression head consists of a fully connected layer that reduces the feature dimension from 1344 to 256, followed by a ReLU activation and a Dropout layer to alleviate overfitting. The final output layer maps the 256-dimensional feature to two target values, corresponding to the Woods--Saxon radius parameter $R_0$ and diffuseness parameter $a$.

The network is trained in a supervised manner using the Huber loss function~\cite{Huber1964RobustEstimation} to mitigate the influence of large residuals. The Woods--Saxon parameters $R_0$ and $a$ are normalized using min-max scaling before training, and the predicted values are inverse-transformed after prediction. The AdamW optimizer is used with an initial learning rate of 0.001, dynamically adjusted using cosine annealing to enhance convergence stability. The model is trained for 100 epochs, with the training set used for optimization, the validation set for monitoring convergence, and the test set for final evaluation. The performance of the model is assessed using the coefficient of determination ($R^2$) and mean absolute error (MAE)~\cite{Kvalseth1985R2,WillmottMatsuura2005MAE}.

\section{Results and Analysis}

In this section, we evaluate the regression performance of the multimodal network for the Woods--Saxon parameters \(R_0\) and \(a\) over eleven centrality selections and four event-group sizes, \(G=1,10,50,\) and \(100\). We first use the inclusive \(0\text{--}100\%\) sample to quantify the improvement obtained from multi-event inputs. We then examine how the regression accuracy varies across ten 10\% centrality intervals and discuss the corresponding geometric interpretation within the TRENTo framework. Complete train-, validation-, and test-set metrics for the individual centrality intervals are provided in Appendix~A.

\subsection{Overall regression performance in the 0--100\% centrality range}

Tables~\ref{tab:r2_group_size} and~\ref{tab:mae_group_size} report the \(R^2\) scores and MAEs for the training, validation, and test sets over the inclusive \(0\text{--}100\%\) centrality range. The three data splits exhibit the same overall dependence on \(G\); the following discussion therefore focuses on the independent test set.

\begin{table}[!htb]
\centering
\caption{$R^2$ scores for the regression of $R_0$ and $a$ under different event-group sizes ($G$) on the training, validation, and test sets (0--100\% centrality).}
\label{tab:r2_group_size}
\begin{tabular}{c|cc|cc|cc}
\hline
\multirow{2}{*}{$G$}
& \multicolumn{2}{c|}{Train}
& \multicolumn{2}{c|}{Validation}
& \multicolumn{2}{c}{Test} \\
& $R^2_{R_0}$ & $R^2_a$ & $R^2_{R_0}$ & $R^2_a$ & $R^2_{R_0}$ & $R^2_a$ \\
\hline
1   & 0.1392 & 0.1156 & 0.1396 & 0.1140 & 0.1349 & 0.1107 \\
10  & 0.6122 & 0.4499 & 0.5671 & 0.3726 & 0.5583 & 0.3706 \\
50  & 0.8175 & 0.5446 & 0.7881 & 0.4396 & 0.7820 & 0.4363 \\
100 & 0.8680 & 0.5770 & 0.8450 & 0.5322 & 0.8389 & 0.5275 \\
\hline
\end{tabular}
\end{table}

\begin{table}[!htb]
\centering
\caption{MAE values (in fm) for the regression of $R_0$ and $a$ under different event-group sizes ($G$) on the training, validation, and test sets (0--100\% centrality).}
\label{tab:mae_group_size}
\begin{tabular}{c|cc|cc|cc}
\hline
\multirow{2}{*}{$G$}
& \multicolumn{2}{c|}{Train}
& \multicolumn{2}{c|}{Validation}
& \multicolumn{2}{c}{Test} \\
& $\mathrm{MAE}_{R_0}$ & $\mathrm{MAE}_a$ & $\mathrm{MAE}_{R_0}$ & $\mathrm{MAE}_a$ & $\mathrm{MAE}_{R_0}$ & $\mathrm{MAE}_a$ \\
\hline
1   & 0.2379 & 0.0485 & 0.2377 & 0.0486 & 0.2328 & 0.0475 \\
10  & 0.1495 & 0.0372 & 0.1584 & 0.0397 & 0.1560 & 0.0387 \\
50  & 0.1039 & 0.0336 & 0.1120 & 0.0371 & 0.1104 & 0.0363 \\
100 & 0.0890 & 0.0328 & 0.0966 & 0.0342 & 0.0960 & 0.0337 \\
\hline
\end{tabular}
\end{table}

With a single event per input, both parameters are only weakly constrained. Increasing the event-group size leads to a clear improvement in the test-set performance. For \(R_0\), \(R^2\) increases from 0.1349 at \(G=1\) to 0.8389 at \(G=100\), while the MAE decreases from 0.2328 to 0.0960~fm. For \(a\), \(R^2\) increases from 0.1107 to 0.5275, and the MAE decreases from 0.0475 to 0.0337~fm over the same range. The results at \(G=10\) and \(50\) lie between these endpoints and follow the same overall trend.

\begin{figure*}[!htb]
  \includegraphics[width=0.9\hsize]{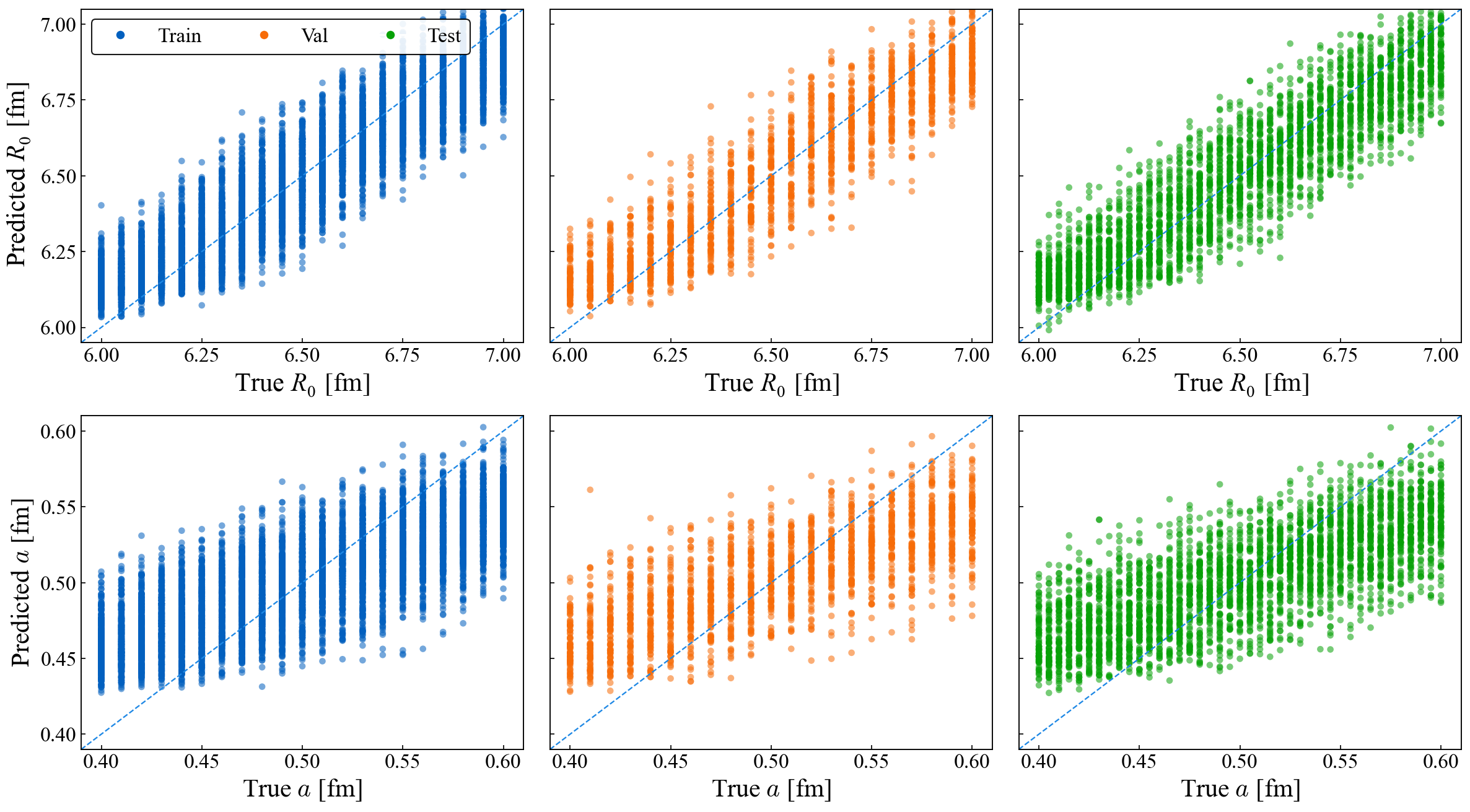}
  \caption{Predicted-versus-true scatter plots of $R_0$ (top row) and $a$ (bottom row) for the training, validation, and test sets in the 0--100\% centrality range with $G=100$. The blue dashed line represents perfect agreement between the predicted and true values.}
  \label{fig:scatter_g100}
\end{figure*}

The quantitative results show that increasing the event-group size improves the regression performance of both \(R_0\) and \(a\), indicating that the multi-event joint-input strategy effectively reduces the influence of fluctuations inherent in single-event inputs. In addition, Fig.~\ref{fig:scatter_g100} shows the predicted-versus-true distributions at \(G=100\). For both parameters, the predicted values generally follow the true values, with the scatter points distributed around the line of perfect agreement. The training, validation, and test sets exhibit similar scatter patterns.

\subsection{Regression performance across centrality ranges and group sizes}

\begin{figure*}[!htb]
  \includegraphics[width=0.9\hsize]{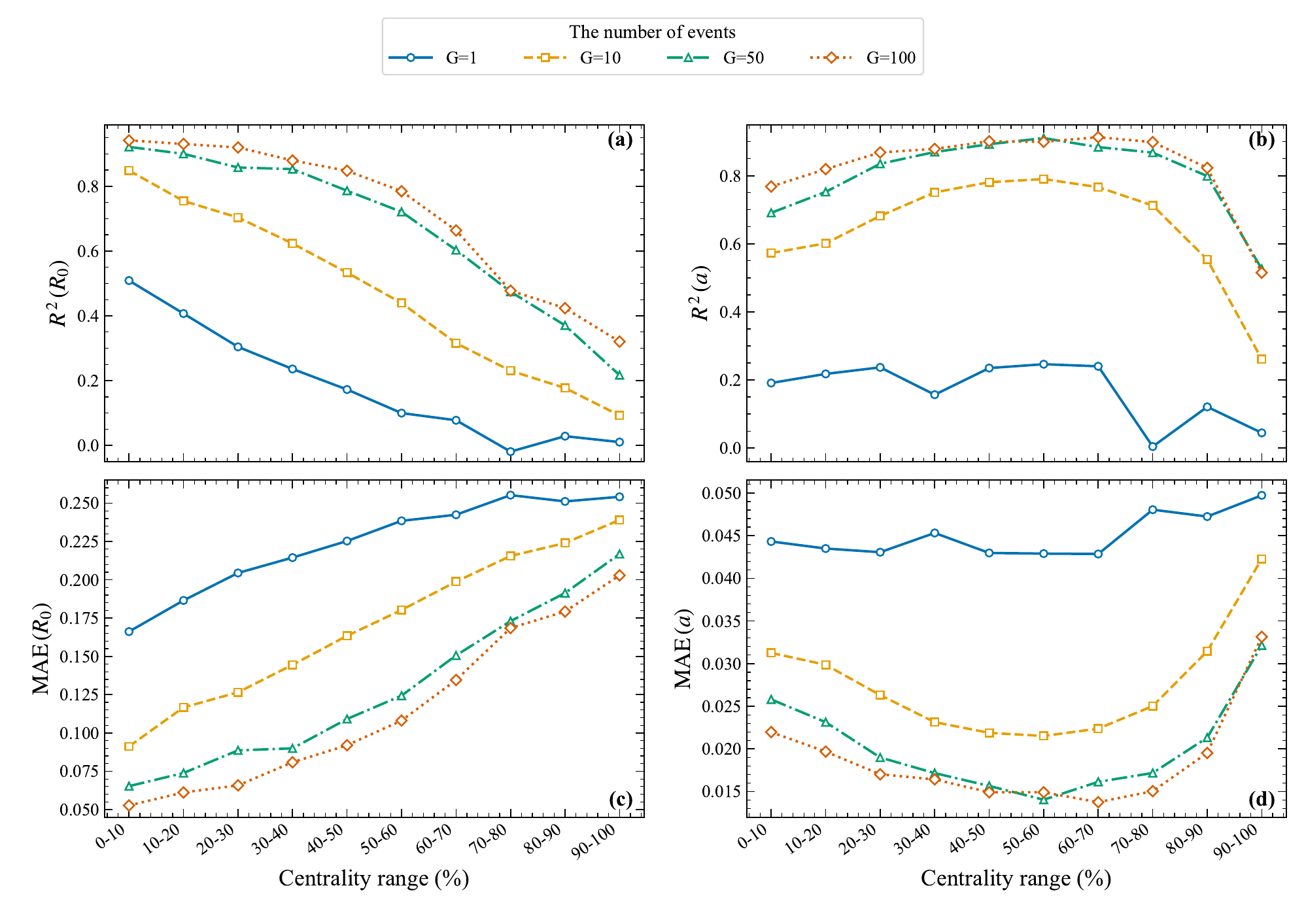}
  \caption{Test-set regression performance for the Woods--Saxon parameters $R_0$ and $a$ as functions of the centrality range for different input event-group sizes ($G=1,\,10,\,50,\,100$). Panels (a) and (b) show the coefficient of determination $R^2$ for $R_0$ and $a$, respectively, while panels (c) and (d) present the corresponding mean absolute error (MAE).}
  \label{fig:tendency}
\end{figure*}

Fig.~\ref{fig:tendency} summarizes the test-set regression performance of \(R_0\) and \(a\) for different centrality intervals and event-group sizes \(G\). Two main trends emerge. First, the regression performance for both parameters generally improves with increasing event-group size \(G\). When \(G\) is further increased from 50 to 100, the improvement in both \(R^2\) and MAE becomes much less pronounced, indicating that the performance gain brought by the multi-event joint-input strategy begins to saturate. This is consistent with the previous subsection. Second, \(R_0\) and \(a\) exhibit markedly different centrality dependences. For the nuclear radius parameter \(R_0\), the regression performance gradually deteriorates from central to peripheral collisions. The best result is obtained in the 0--10\% centrality interval with \(G=100\), where the test-set \(R^2\) reaches 0.9416 and the MAE is reduced to 0.0527~fm. As the collision becomes more peripheral, \(R^2\) decreases continuously, while the MAE increases accordingly. This trend can be understood from the fact that \(R_0\) mainly reflects the overall spatial scale of the nucleus, which is more directly encoded in central collisions where the overlap region is the largest and the global geometry is more robust against fluctuations. By contrast, the diffuseness parameter \(a\) exhibits a clear non-monotonic dependence on centrality, with the best performance observed in the mid-peripheral region. In particular, for the 60--70\% centrality interval with \(G=100\), the test-set \(R^2\) reaches 0.9133 and the MAE decreases to 0.0138~fm, while the performance becomes worse in both more central and more peripheral collisions.

To provide a qualitative geometric interpretation of this behavior, we examine how \(a\) affects the overlap geometry in the TRENTo model. For the Woods--Saxon form above, \(a\) controls the width and steepness of the nuclear surface rather than the position of the half-density surface. Its influence on the reduced thickness function is therefore expected to appear primarily through a redistribution of the overlap profile across its geometric boundary. To illustrate this effect, we consider the reduced thickness function in the TRENTo model, which characterizes the effective overlap of the two incident nuclei and governs the initial entropy deposition in the present setup. It is defined as~\cite{MorelandBernhardBass2015TRENTo}
\begin{equation}
\label{eq:TR}
T_R(x,y) =
\left(
\frac{T_A^p(x,y) + T_B^p(x,y)}{2}
\right)^{1/p},
\end{equation}
where \(T_A(x,y)\) and \(T_B(x,y)\) denote the participant thickness functions of the two incident nuclei, and \(p\) controls the generalized-mean prescription. In the present calculation, \(p=0\) is understood as the continuous \(p\to0\) limit, for which
\begin{equation}
\label{eq:TR1}
T_R(x,y) = \sqrt{T_A(x,y) T_B(x,y)}.
\end{equation}
Changes in \(a\) are therefore reflected in the initial entropy profile through the corresponding variations in \(T_R(x,y)\), which we examine below.

\begin{figure*}[!htb]
  \includegraphics[width=0.9\hsize]{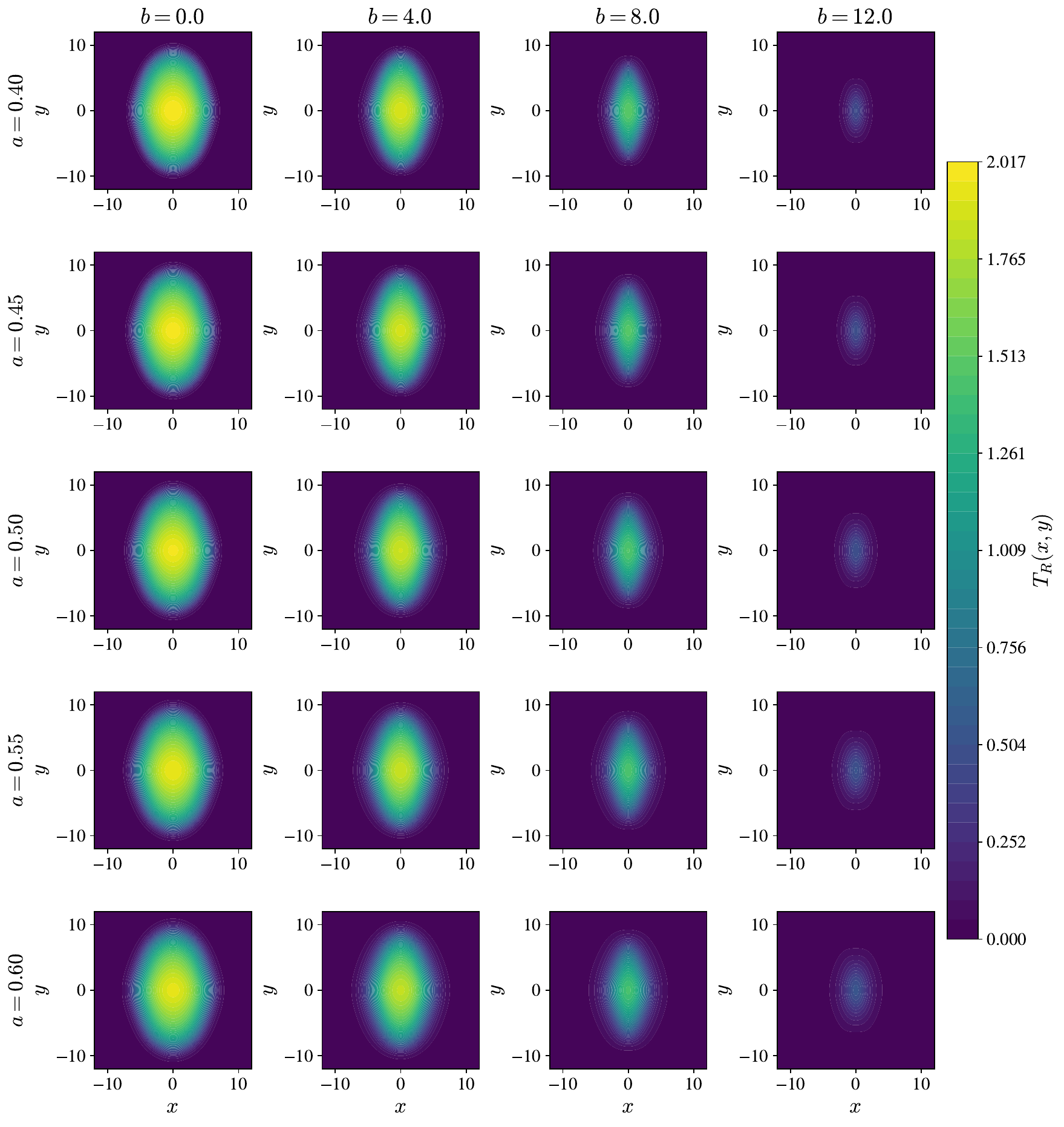}
  \caption{The reduced thickness function \(T_R(x,y)\) for representative body--body configurations at fixed impact parameter \(b\). Each column corresponds to a different value of \(b\), ranging from 0 to 12~fm, and each row corresponds to a different diffuseness parameter \(a\), ranging from 0.40 to 0.60~fm. These fixed-orientation maps are used only to illustrate the geometric effect of \(a\); the regression datasets consist of randomly oriented collisions.}
  \label{fig:overlap}
\end{figure*}

\begin{figure*}[!htb]
  \includegraphics[width=0.9\hsize]{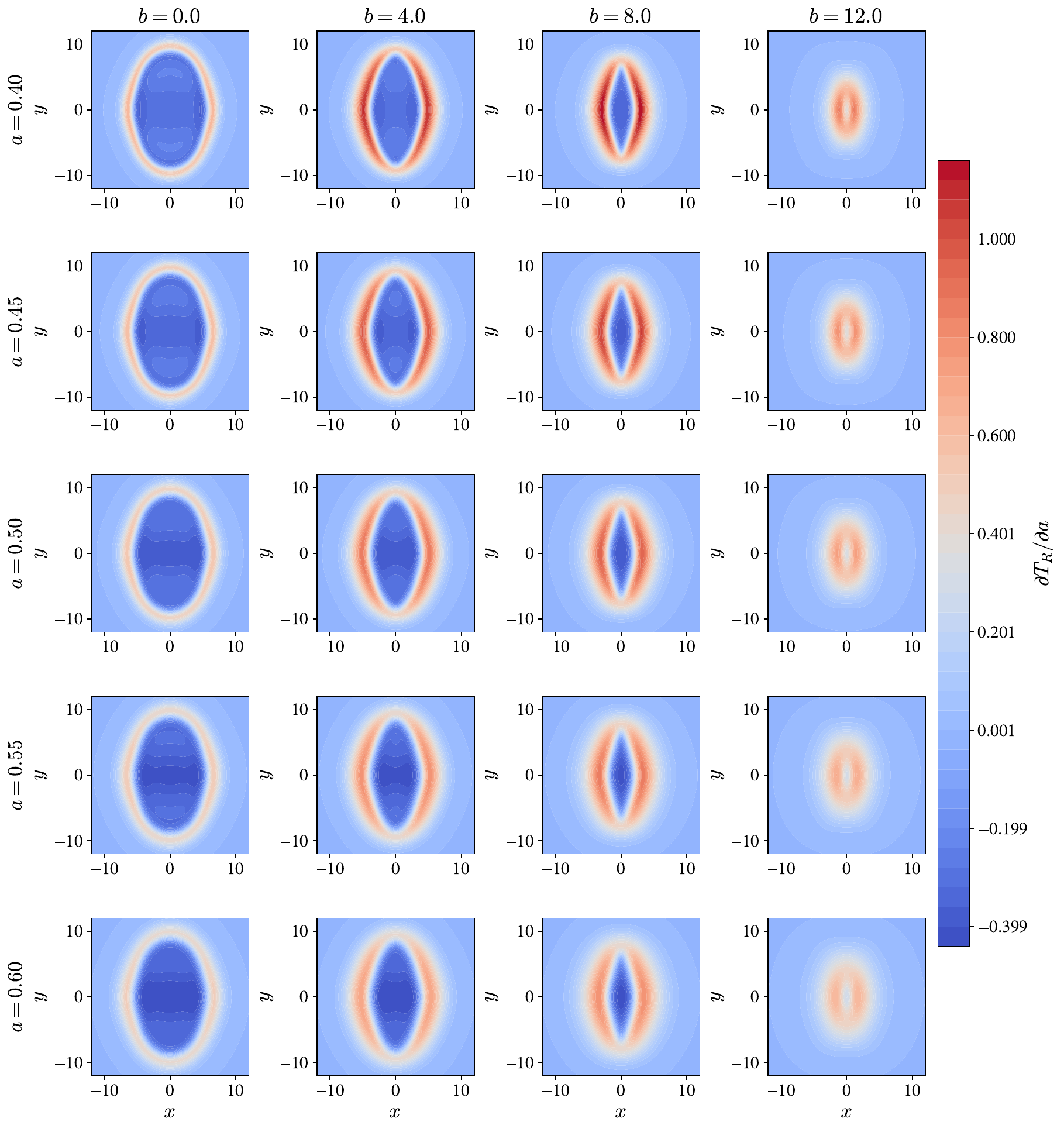}
  \caption{The spatial distribution of \(\partial T_R(x,y)/\partial a\) for representative body--body configurations at fixed impact parameter \(b\). Each column corresponds to a different value of \(b\), ranging from 0 to 12~fm, and each row corresponds to a different value of the diffuseness parameter \(a\), ranging from 0.40 to 0.60~fm.}
  \label{fig:grad}
\end{figure*}

We further investigate the spatial distribution of the reduced thickness function \(T_R(x,y)\) for different values of the diffuseness parameter \(a\) and impact parameter \(b\). Fig.~\ref{fig:overlap} shows the reduced thickness function for body--body collisions under different values of \(a\) and \(b\). For a fixed \(b\), increasing \(a\) enhances the diffuse outer parts of the two nuclear thickness profiles, so the corresponding changes in \(T_R(x,y)\) appear mainly near the edge of the overlap region. As \(b\) increases from head-on to moderately off-central configurations, the overlap becomes more sensitive to these surface contributions. To examine this behavior more directly, we analyze the derivative distribution \(\partial T_R(x,y)/\partial a\). As shown in Fig.~\ref{fig:grad}, the response is concentrated mainly near the boundary of the overlap region and depends non-monotonically on \(b\). It is stronger at \(b=4\) and \(8\)~fm than at \(b=0\) because surface contributions become more important in these moderately off-central configurations, but weakens at \(b=12\)~fm because the overlap is strongly reduced. This behavior is consistent with the higher regression performance in the mid-peripheral region and the lower performance toward the central and peripheral ends. The same behavior is also observed in tip--tip, tip--body, and randomly oriented collisions.

\section{SUMMARY AND OUTLOOK}

In this study, we developed a deep learning framework to invert the Woods--Saxon parameters \(R_0\) and \(a\) from the initial state of high-energy uranium collisions using TRENTo model. The model uses a multimodal fusion network that combines two-dimensional initial-state entropy density images with event-level scalar features such as multiplicity and eccentricities. The image branch employs a pretrained EfficientNet-B1 model to extract deep spatial features from the initial-state images, while the attribute branch processes scalar event features using a 1D convolutional neural network (CNN). The results demonstrate that the initial-state geometry contains information relevant to inferring \(R_0\) and \(a\), with multi-event inputs significantly enhancing the regression accuracy. We also observed that the inversion performance depends on collision centrality, with \(R_0\) being more accurately inferred in central collisions and \(a\) showing the best performance in mid-peripheral collisions.

This study provides a solid foundation for future work in nuclear-structure inversion, offering a method for extracting key nuclear parameters from initial-state information. In the future, we plan to apply this framework to data that has undergone more extensive dynamical evolution, and to explore the extraction of parameters from final-state observables. We will also investigate how these parameters influence the evolution process and the final-state observables, enhancing our understanding of the relationship between nuclear structure and the dynamics of heavy-ion collisions.

\bibliography{}

\appendix
\section*{Appendix}
\section{Regression Results across Group Sizes and Centrality Intervals}

\begin{table*}[!t]
\centering
\caption{$R^2$ scores and MAE values for the regression of $R_0$ and $a$ under different event group sizes $G$ on the training, validation, and test sets (0--10\% centrality).}
\label{tab:r2_mae_0_10}
\begin{tabular}{c|cccc|cccc|cccc}
\hline
\multirow{2}{*}{$G$}
& \multicolumn{4}{c|}{Train}
& \multicolumn{4}{c|}{Validation}
& \multicolumn{4}{c}{Test} \\
& $R^2_{R_0}$ & $\mathrm{MAE}_{R_0}$ & $R^2_a$ & $\mathrm{MAE}_a$
& $R^2_{R_0}$ & $\mathrm{MAE}_{R_0}$ & $R^2_a$ & $\mathrm{MAE}_a$
& $R^2_{R_0}$ & $\mathrm{MAE}_{R_0}$ & $R^2_a$ & $\mathrm{MAE}_a$ \\
\hline
1
& 0.5560 & 0.1625 & 0.2600 & 0.0435
& 0.5197 & 0.1687 & 0.1945 & 0.0453
& 0.5089 & 0.1662 & 0.1911 & 0.0443 \\
10
& 0.8651 & 0.0883 & 0.6196 & 0.0304
& 0.8555 & 0.0911 & 0.5810 & 0.0317
& 0.8485 & 0.0912 & 0.5729 & 0.0313 \\
50
& 0.9781 & 0.0360 & 0.9745 & 0.0077
& 0.9293 & 0.0637 & 0.7036 & 0.0258
& 0.9214 & 0.0653 & 0.6910 & 0.0258 \\
100
& 0.9812 & 0.0327 & 0.9514 & 0.0106
& 0.9493 & 0.0533 & 0.7913 & 0.0217
& 0.9416 & 0.0527 & 0.7681 & 0.0220 \\
\hline
\end{tabular}
\end{table*}

\begin{table*}[!t]
\centering
\caption{$R^2$ scores and MAE values for the regression of $R_0$ and $a$ under different event group sizes $G$ on the training, validation, and test sets (10--20\% centrality).}
\label{tab:r2_mae_10_20}
\begin{tabular}{c|cccc|cccc|cccc}
\hline
\multirow{2}{*}{$G$}
& \multicolumn{4}{c|}{Train}
& \multicolumn{4}{c|}{Validation}
& \multicolumn{4}{c}{Test} \\
& $R^2_{R_0}$ & $\mathrm{MAE}_{R_0}$ & $R^2_a$ & $\mathrm{MAE}_a$
& $R^2_{R_0}$ & $\mathrm{MAE}_{R_0}$ & $R^2_a$ & $\mathrm{MAE}_a$
& $R^2_{R_0}$ & $\mathrm{MAE}_{R_0}$ & $R^2_a$ & $\mathrm{MAE}_a$ \\
\hline
1
& 0.4273 & 0.1881 & 0.2355 & 0.0440
& 0.4197 & 0.1895 & 0.2311 & 0.0442
& 0.4071 & 0.1865 & 0.2178 & 0.0435 \\
10
& 0.7739 & 0.1153 & 0.6526 & 0.0287
& 0.7597 & 0.1185 & 0.6271 & 0.0297
& 0.7544 & 0.1167 & 0.6013 & 0.0299 \\
50
& 0.9660 & 0.0447 & 0.9480 & 0.0111
& 0.9008 & 0.0753 & 0.7681 & 0.0230
& 0.9000 & 0.0739 & 0.7526 & 0.0231 \\
100
& 0.9726 & 0.0401 & 0.9401 & 0.0119
& 0.9326 & 0.0620 & 0.8261 & 0.0198
& 0.9304 & 0.0612 & 0.8193 & 0.0197 \\
\hline
\end{tabular}
\end{table*}

\begin{table*}[!t]
\centering
\caption{$R^2$ scores and MAE values for the regression of $R_0$ and $a$ under different event group sizes $G$ on the training, validation, and test sets (20--30\% centrality).}
\label{tab:r2_mae_20_30}
\begin{tabular}{c|cccc|cccc|cccc}
\hline
\multirow{2}{*}{$G$}
& \multicolumn{4}{c|}{Train}
& \multicolumn{4}{c|}{Validation}
& \multicolumn{4}{c}{Test} \\
& $R^2_{R_0}$ & $\mathrm{MAE}_{R_0}$ & $R^2_a$ & $\mathrm{MAE}_a$
& $R^2_{R_0}$ & $\mathrm{MAE}_{R_0}$ & $R^2_a$ & $\mathrm{MAE}_a$
& $R^2_{R_0}$ & $\mathrm{MAE}_{R_0}$ & $R^2_a$ & $\mathrm{MAE}_a$ \\
\hline
1
& 0.3125 & 0.2084 & 0.2467 & 0.0438
& 0.3116 & 0.2085 & 0.2442 & 0.0439
& 0.3043 & 0.2045 & 0.2370 & 0.0431 \\
10
& 0.7371 & 0.1230 & 0.7271 & 0.0251
& 0.7122 & 0.1281 & 0.6993 & 0.0263
& 0.7035 & 0.1265 & 0.6819 & 0.0263 \\
50
& 0.9053 & 0.0746 & 0.9082 & 0.0147
& 0.8681 & 0.0874 & 0.8578 & 0.0179
& 0.8578 & 0.0886 & 0.8350 & 0.0190 \\
100
& 0.9534 & 0.0518 & 0.9270 & 0.0132
& 0.9203 & 0.0676 & 0.8794 & 0.0167
& 0.9197 & 0.0658 & 0.8684 & 0.0170 \\
\hline
\end{tabular}
\end{table*}

\begin{table*}[!t]
\centering
\caption{$R^2$ scores and MAE values for the regression of $R_0$ and $a$ under different event group sizes $G$ on the training, validation, and test sets (30--40\% centrality).}
\label{tab:r2_mae_30_40}
\begin{tabular}{c|cccc|cccc|cccc}
\hline
\multirow{2}{*}{$G$}
& \multicolumn{4}{c|}{Train}
& \multicolumn{4}{c|}{Validation}
& \multicolumn{4}{c}{Test} \\
& $R^2_{R_0}$ & $\mathrm{MAE}_{R_0}$ & $R^2_a$ & $\mathrm{MAE}_a$
& $R^2_{R_0}$ & $\mathrm{MAE}_{R_0}$ & $R^2_a$ & $\mathrm{MAE}_a$
& $R^2_{R_0}$ & $\mathrm{MAE}_{R_0}$ & $R^2_a$ & $\mathrm{MAE}_a$ \\
\hline
1
& 0.2516 & 0.2177 & 0.1653 & 0.0462
& 0.2478 & 0.2182 & 0.1633 & 0.0462
& 0.2360 & 0.2145 & 0.1567 & 0.0453 \\
10
& 0.6815 & 0.1368 & 0.7899 & 0.0219
& 0.6362 & 0.1455 & 0.7613 & 0.0232
& 0.6233 & 0.1445 & 0.7510 & 0.0231 \\
50
& 0.8792 & 0.0838 & 0.8850 & 0.0165
& 0.8670 & 0.0880 & 0.8717 & 0.0174
& 0.8529 & 0.0900 & 0.8696 & 0.0172 \\
100
& 0.9635 & 0.0464 & 0.9429 & 0.0116
& 0.8937 & 0.0770 & 0.8863 & 0.0164
& 0.8795 & 0.0809 & 0.8788 & 0.0164 \\
\hline
\end{tabular}
\end{table*}

\begin{table*}[!t]
\centering
\caption{$R^2$ scores and MAE values for the regression of $R_0$ and $a$ under different event group sizes $G$ on the training, validation, and test sets (40--50\% centrality).}
\label{tab:r2_mae_40_50}
\begin{tabular}{c|cccc|cccc|cccc}
\hline
\multirow{2}{*}{$G$}
& \multicolumn{4}{c|}{Train}
& \multicolumn{4}{c|}{Validation}
& \multicolumn{4}{c}{Test} \\
& $R^2_{R_0}$ & $\mathrm{MAE}_{R_0}$ & $R^2_a$ & $\mathrm{MAE}_a$
& $R^2_{R_0}$ & $\mathrm{MAE}_{R_0}$ & $R^2_a$ & $\mathrm{MAE}_a$
& $R^2_{R_0}$ & $\mathrm{MAE}_{R_0}$ & $R^2_a$ & $\mathrm{MAE}_a$ \\
\hline
1
& 0.1877 & 0.2288 & 0.2493 & 0.0436
& 0.1802 & 0.2299 & 0.2461 & 0.0437
& 0.1727 & 0.2253 & 0.2350 & 0.0430 \\
10
& 0.5916 & 0.1569 & 0.8033 & 0.0213
& 0.5503 & 0.1643 & 0.7878 & 0.0220
& 0.5334 & 0.1634 & 0.7809 & 0.0219 \\
50
& 0.8457 & 0.0968 & 0.9033 & 0.0152
& 0.7953 & 0.1102 & 0.8982 & 0.0155
& 0.7861 & 0.1091 & 0.8926 & 0.0157 \\
100
& 0.8844 & 0.0822 & 0.9104 & 0.0146
& 0.8622 & 0.0889 & 0.9045 & 0.0152
& 0.8477 & 0.0920 & 0.9010 & 0.0149 \\
\hline
\end{tabular}
\end{table*}

\begin{table*}[!t]
\centering
\caption{$R^2$ scores and MAE values for the regression of $R_0$ and $a$ under different event group sizes $G$ on the training, validation, and test sets (50--60\% centrality).}
\label{tab:r2_mae_50_60}
\begin{tabular}{c|cccc|cccc|cccc}
\hline
\multirow{2}{*}{$G$}
& \multicolumn{4}{c|}{Train}
& \multicolumn{4}{c|}{Validation}
& \multicolumn{4}{c}{Test} \\
& $R^2_{R_0}$ & $\mathrm{MAE}_{R_0}$ & $R^2_a$ & $\mathrm{MAE}_a$
& $R^2_{R_0}$ & $\mathrm{MAE}_{R_0}$ & $R^2_a$ & $\mathrm{MAE}_a$
& $R^2_{R_0}$ & $\mathrm{MAE}_{R_0}$ & $R^2_a$ & $\mathrm{MAE}_a$ \\
\hline
1
& 0.1116 & 0.2427 & 0.2569 & 0.0436
& 0.1040 & 0.2439 & 0.2524 & 0.0437
& 0.0999 & 0.2385 & 0.2463 & 0.0429 \\
10
& 0.4884 & 0.1769 & 0.8105 & 0.0210
& 0.4487 & 0.1838 & 0.8034 & 0.0213
& 0.4397 & 0.1803 & 0.7902 & 0.0215 \\
50
& 0.7802 & 0.1150 & 0.9190 & 0.0137
& 0.7219 & 0.1272 & 0.9161 & 0.0139
& 0.7208 & 0.1244 & 0.9105 & 0.0141 \\
100
& 0.8572 & 0.0918 & 0.9055 & 0.0148
& 0.7934 & 0.1078 & 0.9070 & 0.0147
& 0.7846 & 0.1081 & 0.8995 & 0.0149 \\
\hline
\end{tabular}
\end{table*}

\begin{table*}[!t]
\centering
\caption{$R^2$ scores and MAE values for the regression of $R_0$ and $a$ under different event group sizes $G$ on the training, validation, and test sets (60--70\% centrality).}
\label{tab:r2_mae_60_70}
\begin{tabular}{c|cccc|cccc|cccc}
\hline
\multirow{2}{*}{$G$}
& \multicolumn{4}{c|}{Train}
& \multicolumn{4}{c|}{Validation}
& \multicolumn{4}{c}{Test} \\
& $R^2_{R_0}$ & $\mathrm{MAE}_{R_0}$ & $R^2_a$ & $\mathrm{MAE}_a$
& $R^2_{R_0}$ & $\mathrm{MAE}_{R_0}$ & $R^2_a$ & $\mathrm{MAE}_a$
& $R^2_{R_0}$ & $\mathrm{MAE}_{R_0}$ & $R^2_a$ & $\mathrm{MAE}_a$ \\
\hline
1
& 0.1140 & 0.2434 & 0.2672 & 0.0431
& 0.0820 & 0.2477 & 0.2463 & 0.0437
& 0.0778 & 0.2424 & 0.2402 & 0.0429 \\
10
& 0.3929 & 0.1925 & 0.7897 & 0.0219
& 0.3466 & 0.1995 & 0.7753 & 0.0225
& 0.3158 & 0.1988 & 0.7664 & 0.0224 \\
50
& 0.6959 & 0.1366 & 0.8967 & 0.0158
& 0.6224 & 0.1522 & 0.8908 & 0.0163
& 0.6032 & 0.1506 & 0.8841 & 0.0161 \\
100
& 0.8652 & 0.0896 & 0.9267 & 0.0131
& 0.6827 & 0.1357 & 0.9198 & 0.0138
& 0.6637 & 0.1346 & 0.9133 & 0.0138 \\
\hline
\end{tabular}
\end{table*}

\begin{table*}[!t]
\centering
\caption{$R^2$ scores and MAE values for the regression of $R_0$ and $a$ under different event group sizes $G$ on the training, validation, and test sets (70--80\% centrality).}
\label{tab:r2_mae_70_80}
\begin{tabular}{c|cccc|cccc|cccc}
\hline
\multirow{2}{*}{$G$}
& \multicolumn{4}{c|}{Train}
& \multicolumn{4}{c|}{Validation}
& \multicolumn{4}{c}{Test} \\
& $R^2_{R_0}$ & $\mathrm{MAE}_{R_0}$ & $R^2_a$ & $\mathrm{MAE}_a$
& $R^2_{R_0}$ & $\mathrm{MAE}_{R_0}$ & $R^2_a$ & $\mathrm{MAE}_a$
& $R^2_{R_0}$ & $\mathrm{MAE}_{R_0}$ & $R^2_a$ & $\mathrm{MAE}_a$ \\
\hline
1
& -0.0153 & 0.2610 & 0.0223 & 0.0488
& -0.0157 & 0.2610 & 0.0185 & 0.0488
& -0.0190 & 0.2553 & 0.0044 & 0.0480 \\
10
& 0.2690 & 0.2155 & 0.7373 & 0.0246
& 0.2449 & 0.2186 & 0.7192 & 0.0254
& 0.2306 & 0.2155 & 0.7126 & 0.0250 \\
50
& 0.6161 & 0.1531 & 0.8824 & 0.0165
& 0.4854 & 0.1765 & 0.8684 & 0.0175
& 0.4745 & 0.1730 & 0.8679 & 0.0172 \\
100
& 0.8137 & 0.1059 & 0.9186 & 0.0140
& 0.5057 & 0.1670 & 0.9030 & 0.0152
& 0.4778 & 0.1685 & 0.8987 & 0.0150 \\
\hline
\end{tabular}
\end{table*}

\begin{table*}[!t]
\centering
\caption{$R^2$ scores and MAE values for the regression of $R_0$ and $a$ under different event group sizes $G$ on the training, validation, and test sets (80--90\% centrality).}
\label{tab:r2_mae_80_90}
\begin{tabular}{c|cccc|cccc|cccc}
\hline
\multirow{2}{*}{$G$}
& \multicolumn{4}{c|}{Train}
& \multicolumn{4}{c|}{Validation}
& \multicolumn{4}{c}{Test} \\
& $R^2_{R_0}$ & $\mathrm{MAE}_{R_0}$ & $R^2_a$ & $\mathrm{MAE}_a$
& $R^2_{R_0}$ & $\mathrm{MAE}_{R_0}$ & $R^2_a$ & $\mathrm{MAE}_a$
& $R^2_{R_0}$ & $\mathrm{MAE}_{R_0}$ & $R^2_a$ & $\mathrm{MAE}_a$ \\
\hline
1
& 0.0322 & 0.2567 & 0.1243 & 0.0483
& 0.0302 & 0.2569 & 0.1226 & 0.0483
& 0.0287 & 0.2512 & 0.1210 & 0.0472 \\
10
& 0.2262 & 0.2227 & 0.5874 & 0.0310
& 0.1868 & 0.2281 & 0.5610 & 0.0320
& 0.1773 & 0.2240 & 0.5537 & 0.0315 \\
50
& 0.5165 & 0.1735 & 0.8347 & 0.0200
& 0.3772 & 0.1969 & 0.8052 & 0.0216
& 0.3705 & 0.1914 & 0.7988 & 0.0213 \\
100
& 0.7052 & 0.1338 & 0.8641 & 0.0180
& 0.3949 & 0.1908 & 0.8237 & 0.0203
& 0.4238 & 0.1793 & 0.8229 & 0.0195 \\
\hline
\end{tabular}
\end{table*}

\begin{table*}[!t]
\centering
\caption{$R^2$ scores and MAE values for the regression of $R_0$ and $a$ under different event group sizes $G$ on the training, validation, and test sets (90--100\% centrality).}
\label{tab:r2_mae_90_100}
\begin{tabular}{c|cccc|cccc|cccc}
\hline
\multirow{2}{*}{$G$}
& \multicolumn{4}{c|}{Train}
& \multicolumn{4}{c|}{Validation}
& \multicolumn{4}{c}{Test} \\
& $R^2_{R_0}$ & $\mathrm{MAE}_{R_0}$ & $R^2_a$ & $\mathrm{MAE}_a$
& $R^2_{R_0}$ & $\mathrm{MAE}_{R_0}$ & $R^2_a$ & $\mathrm{MAE}_a$
& $R^2_{R_0}$ & $\mathrm{MAE}_{R_0}$ & $R^2_a$ & $\mathrm{MAE}_a$ \\
\hline
1
& 0.0139 & 0.2598 & 0.0464 & 0.0509
& 0.0139 & 0.2598 & 0.0442 & 0.0509
& 0.0105 & 0.2542 & 0.0447 & 0.0497 \\
10
& 0.1625 & 0.2352 & 0.3211 & 0.0415
& 0.1150 & 0.2419 & 0.2612 & 0.0432
& 0.0927 & 0.2391 & 0.2611 & 0.0423 \\
50
& 0.2921 & 0.2106 & 0.5755 & 0.0313
& 0.2592 & 0.2162 & 0.5154 & 0.0332
& 0.2171 & 0.2168 & 0.5260 & 0.0321 \\
100
& 0.4533 & 0.1862 & 0.5807 & 0.0316
& 0.3814 & 0.1969 & 0.5193 & 0.0336
& 0.3205 & 0.2029 & 0.5152 & 0.0331 \\
\hline
\end{tabular}
\end{table*}

\end{document}